\documentclass[fleqn,12pt,twoside]{article}
\usepackage[headings]{espcrc1}

\readRCS
$Id: espcrc1.tex,v 1.2 2004/02/24 11:22:11 spepping Exp $
\usepackage{graphicx}
\usepackage[figuresright]{rotating}

\newcommand{\AmS}{{\protect\the\textfont2
  A\kern-.1667em\lower.5ex\hbox{M}\kern-.125emS}}

\title{Properties of stellar matter in supernova explosions and nuclear 
multifragmentation} 

\author{A.S. Botvina\address[INR]{Institute for Nuclear Research, 
Russian Academy of Sciences, 117312 Moscow, Russia} 
\address[GSI]{Gesellschaft f{\"u}r Schwerionenforschung 
mbH, D-64291 Darmstadt, Germany} \address[FIAS]
{Frankfurt Institute for Advanced Studies, J.W. Goethe University, 
D-60438 Frankfurt am Main, Germany}, 
I.N. Mishustin\addressmark[FIAS] \address[KUR]{Kurchatov Institute, Russian 
Research Center, 123182 Moscow, Russia}, 
and W. Trautmann\addressmark[GSI]}

\runtitle{Properties of stellar matter in supernova explosions and nuclear 
multifragmentation}
\runauthor{A.S. Botvina, I.N. Mishustin and W.Trautmann}

\begin{document}

\maketitle






\begin{abstract}
During the collapse of massive stars, and the supernova type-II 
explosions, stellar matter reaches densities 
and temperatures which are similar to the ones obtained in
intermediate-energy nucleus-nucleus collisions. The 
nuclear multifragmentation reactions can be used for determination
of properties of nuclear matter at subnuclear densities,
in the region of the nuclear liquid-gas phase transition. It 
is demonstrated that the modified properties of hot nuclei (in particular,
their symmetry energy) extracted from the multifragmentation data
can essentially influence nuclear composition of stellar matter.
The effects on weak processes, and on the nucleosynthesis are 
also discussed. 
\end{abstract}



\section{INTRODUCTION}

One of the most spectacular events in astrophysics is the type II supernova 
explosion which releases of about $10^{53}$ erg of energy, or several
tens of MeV per nucleon \cite{Bethe}.
When the core of a massive star collapses, it reaches densities several
times larger than the normal nuclear density $\rho_0\approx 0.15$ fm$^{-3}$.
The repulsive nucleon-nucleon interaction gives rise to a bounce-off
and creation of a shock wave propagating through the in-falling stellar
material. This shock wave is responsible for the ejection of a star 
envelope that is observed as a supernova explosion.
During the collapse and subsequent explosion the temperatures
$T\approx (0.5\div 10)$ MeV and baryon densities $\rho \approx
(10^{-5}\div 2) \rho_0$ can be reached. As shown by many
theoretical studies, a liquid-gas phase transition is expected in
nuclear matter under these conditions. It is remarkable that
similar conditions can be obtained in energetic nuclear collisions 
in terrestrial laboratories, which lead to multifragmentation 
reactions. 

Multifragmentation, i.e. a break-up of nuclei into many small fragments, 
has been observed in nearly all types of nuclear reactions when a large 
amount of energy is deposited in nuclei. It includes reactions 
induced by protons, pions, antiprotons, and by heavy ions of both, 
relativistic energies (peripheral collisions) and 
'Fermi'-energies (central collisions) 
\cite{SMM,Botvina95,EOS,ISIS,MSU,INDRA,FAZA,Xi}.
According to the present understanding, multifragmentation is a relatively 
fast process, 
with a characteristic time around 100 fm/c, where, nevertheless, a high
degree of equilibration is reached. The process is mainly
associated with abundant production of intermediate mass fragments
(IMFs, with charges $Z \approx$ 3-20). However, at the onset of
multifragmentation, also heavy residues are produced which have  
previously only been associated with compound-nucleus processes. At
very high excitation energies, the IMF production gives way
to the total vaporization of nuclei into nucleons and very light clusters.

\section{PHASE DIAGRAM OF NUCLEAR MATTER}

\vspace{-10mm}
\begin{figure}[htb]
\centering
\begin{minipage}[t]{80mm}
\centerline{\includegraphics[height=9.5cm]{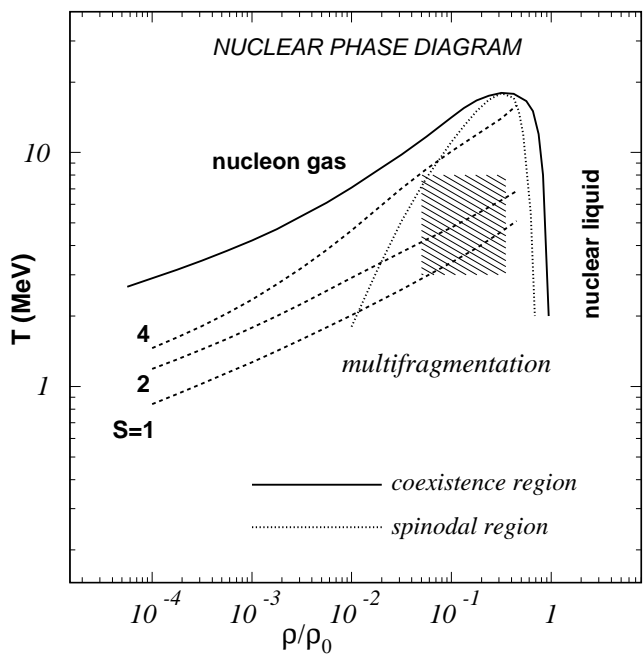}}
\end{minipage}
\vspace{-5mm}
\caption{Nuclear phase diagram on the temperature--density 
plane. Solid and dotted lines give borders of the liquid-gas 
coexistence region and the spinodal region. The shaded 
area corresponds to conditions reached in nuclear multifragmentation 
reactions. The dashed lines are isentropic trajectories characterized 
by constant entropies per baryon ($S=$1, 2, and 4).}
\end{figure}

\vspace{-10mm}

The multifragmentation reaction can be considered as an experimental tool 
to study the properties of hot fragments and the phase diagram of nuclear 
matter at densities $\rho \sim 0.1 \rho_0$ and temperatures around 
$T \approx$ 3--8 MeV which are expected to be reached in the
freeze-out volume. In Fig.~1 we demonstrate a schematic phase diagram of 
nuclear matter which has a liquid-gas phase transition. The shaded area 
indicates the region of densities and temperatures which can be studied 
in nuclear multifragmentation processes. We have also shown isentropic 
trajectories with S/B values of 1, 2, and 4 typical for supernova 
explosions. One can see, for example, that a nearly 
adiabatic collapse of the massive stars with typical entropies 
of 1-2 per baryon passes exactly through the multifragmentation 
area. 

\section{IN-MEDIUM MODIFICATION OF NUCLEAR PROPERTIES}

Multifragmentation opens a unique possibility to
investigate this part of the phase diagram. In particular, 
the "in-medium" modifications of properties of hot nuclei are very 
important for astrophysical applications \cite{Botvina04,Botvina05}. 
Recently, the symmetry energy of hot nuclei was extracted in
ref. \cite{LeFevre}, and it was demonstrated that it decreased 
considerably from the values expected for cold isolated nuclei 
with increasing excitation energy at multifragmentation. 
In Fig.~2 we show the symmetry energy coefficient $\gamma$, which becomes 
lower with decrease of the impact parameters $b$, i.e. for the 
more central collisions. In this case 
the symmetry energy of hot fragments with mass
number $A$ and charge $Z$ is defined as
$E^{\rm sym}_{A,Z}=\gamma (A-2Z)^2/A$. The 
phenomenological parameter $\gamma$ is approximately $25$~MeV 
for isolated nuclei, in order to describe their binding energies. 
As one see it decreases down to $\approx 15$ MeV, and it may be even 
lower, as shown in the analysis \cite{LeFevre}. 

\vspace{-8mm}
\begin{figure}[htb]
\centering
\begin{minipage}[t]{80mm}
\centerline{\includegraphics[height=6.2cm]{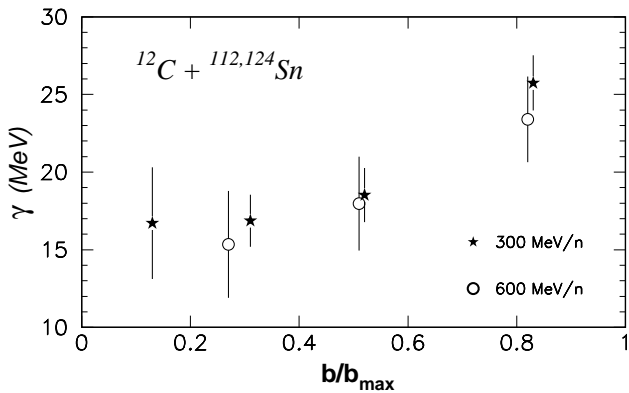}}
\end{minipage}
\vspace{-6mm}
\caption{The apparent symmetry energy coefficient 
$\gamma$ of hot nuclei, as extracted from multifragmentation of tin 
isotopes induced by $^{12}C$ beams with energy 300 and 600 
MeV per nucleon, versus relative impact parameter $b/b_{max}$ 
\cite{LeFevre}.}
\end{figure}

\vspace{-7mm}

As a model which can provide connection between nuclear multifragmentation 
and astrophysical processes we take the Statistical
Multifragmentation Model (SMM), for a review see ref. \cite{SMM}. 
As demonstrated by many analyses  
\cite{Botvina95,EOS,ISIS,MSU,INDRA,FAZA,Xi}, the model describes 
experimental data very well. 
Here a reduction of the symmetry term can be considered as a result of 
modification of the fragments in hot environment, including mutual 
interactions between them. 
Since the SMM can be applied both for finite systems and in
the thermodynamical limit for infinite systems, it may be generalized 
for supernova conditions, where a nuclear statistical equilibration is 
usually expected. This generalization was performed in 
refs.\cite{Botvina04,Botvina05} by including effects of electron, neutrino, 
and photon interactions in stellar matter.

\section{ASTROPHYSICAL IMPORTANCE OF NUCLEAR PROPERTIES}

In Fig.~3 we demonstrate the results of SMM calculations both 
for multifragmentation of $Au$ sources at different excitation energies, 
and for a stellar matter with density, electron fraction, and temperatures 
expected during the collapse of massive stars and supernova explosions. 
One can see that the evolution of mass distributions with excitation 
energy is qualitatively the same for both the nuclear multifragmentation 
reactions and the supernova process. The transition with excitation 
energy (and with temperature) from the 'U-shaped' mass distribution to 
the exponential distribution, a characteristic feature of the liquid-gas 
phase transition, is taking place in both cases. 
However, in the supernova environments much heavier and neutron-rich 
nuclei can be produced because of screening effect of 
surrounding electrons. 

\vspace{-6mm}
\begin{figure}[htb]
\centering
\begin{minipage}[t]{80mm}
\centerline{\includegraphics[height=10.9cm]{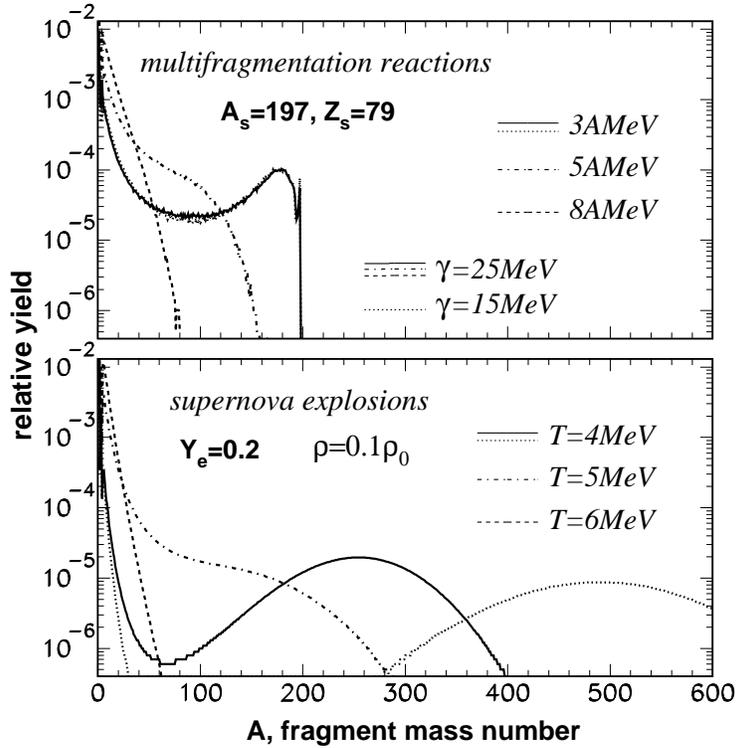}}
\end{minipage}
\vspace{-6mm}
\caption{Relative yields of fragments (per one nucleon) in 
multifragmentation of $Au$ sources (top panel) and 
in supernova environment at the electron fraction $Y_e=$0.2 per 
nucleon and the density of 0.1$\rho_0$ (bottom panel). The 
calculations at excitation energies of 3, 5, and 8 MeV per nucleon (top), 
and different temperatures $T$ (bottom), are shown by different curves.
Effects of the reduced symmetry energy coefficients $\gamma$ 
are also demonstrated (top and bottom).}
\end{figure}

In Fig.~3 we have also shown how important is the information about the 
symmetry energy of hot nuclei extracted from the experiment. 
One can see from mass yields at 3 MeV per nucleon in top panel, 
changing $\gamma$ coefficient from 25 to 15 MeV has practically no 
influence on mean mass distributions of fragments produced in 
nuclear reactions. As was discussed, in the case of multifragmentation 
of finite nuclei the isotope distributions become wider at smaller 
$\gamma$ \cite{Botvina06,nihal}. However, the $\gamma$ has a dramatic 
influence on masses of nuclei produced in supernova environment. 
It is seen that a lot of superheavy (and exotic) nuclei are produced 
in this case. Their production will influence dynamics of the collapse 
and explosion. In the following these hot nuclei should undergo 
de-excitation, and their decay products can serve as seeds for subsequent 
$r-$ process. 
In this respect, studying the multifragmentation reactions in the 
laboratory is important for understanding how heavy elements were 
synthesized in the Universe.

Changing the symmetry energy of nuclei is also very important 
for weak interactions. In Fig.~4 we demonstrate that the electron 
capture rate in stellar matter depends essentially on its value. 
The calculations of this rate $R_e$ (per nucleon and 
per second) was carried out with the method suggested in ref. 
\cite{langanke}, which is based on an independent particle model 
and dominance of Gamow-Teller transitions. One can see that at 
relatively high densities $\rho \sim 0.1\rho_0$ the electron 
capture rate changes only by 20-50\%, if we adopt 
the reduced symmetry energy coefficient. This is because a high 
electron chemical potential drives the reaction. 
However, at small densities, when large nuclei still exist (i.e. at 
low temperatures), the effect of $\gamma$ could be dramatic, by two-three 
orders of magnitude. We note, that at these relatively small densities and 
temperatures the nuclear chemical equilibrium is usually assumed to be 
established \cite{network}. We believe that hot nuclei interact 
with each other by neutron exchange in this case. This situation 
is similar to what we have at higher densities of nuclear matter 
in multifragmentation reactions. Therefore, the effect of reduction 
of the symmetry energy observed in multifragmentation may also take 
place at small densities in supernova environment.

\vspace{-7mm}
\begin{figure}[htb]
\centering
\begin{minipage}[t]{80mm}
\centerline{\includegraphics[height=9.5cm]{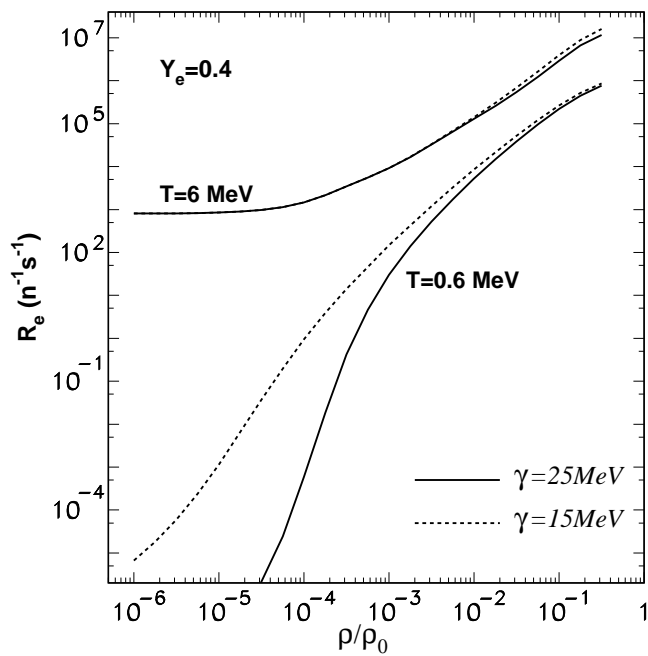}}
\end{minipage}
\vspace{-6mm}
\caption{Density dependence of electron-capture rates $R_e$ 
on hot nuclei in supernova environment at different 
temperatures $T$ and the electron fraction $Y_e$=0.4.
Solid and dashed lines show results for standard (25 MeV) and 
reduced (15 MeV) values of the symmetry energy coefficients $\gamma$.}
\end{figure}

\section{CONCLUSIONS}

We have pointed out that similar physical conditions of 
nuclear matter are reached in multifragmentation reactions and in 
explosion of massive stars. Statistical models successfully applied 
for description of nuclear multifragmentation can be used for 
astrophysical conditions too. Input parameters of the 
models (e.g. their symmetry energy, surface energy) can be directly 
extracted from multifragmentation experiments. Broad variety of nuclei 
including exotic and neutron-rich ones are produced in stellar matter. 
Modification of the symmetry energy observed in hot nuclei in dense 
environment may have strong impact on the weak reaction rates 
and nucleosynthesis of heavy elements.

\end{document}